\documentclass[twocolumn,prb,aps,superscriptaddress,longbibliography]{revtex4-2}
\usepackage{amsmath}
\usepackage{amssymb}
\usepackage{amsfonts}
\usepackage[dvips]{graphicx}
\usepackage{subfigure}
\usepackage{dcolumn}
\usepackage{txfonts}
\usepackage{bm}
\usepackage{makeidx}
\usepackage{color}
\usepackage{mathtools}
\usepackage{threeparttable}
\usepackage{physics}
\usepackage[colorlinks,linkcolor=blue,anchorcolor=blue,citecolor=blue,urlcolor=blue]{hyperref}
\usepackage{lipsum}

\begin{document}
\title{Feynman Paradox about the Josephson effect and a sawtooth current in the double junction}

\author{Zhi-Lei Zhang}
\affiliation{Graduate School of China Academy of Engineering Physics, Beijing 100193, China}

\author{Guo-Jian Qiao}
\affiliation{Beijing Computational Science Research Center, Beijing 100193, China}

\author{C. P. Sun}
\email{suncp@gscaep.ac.cn}
\affiliation{Graduate School of China Academy of Engineering Physics, Beijing 100193, China}
\affiliation{Beijing Computational Science Research Center, Beijing 100193, China}
 
\begin{abstract}
We revisit the Feynman approach to the Josephson effect, which employs a pair of linear coupling equations for its modeling. It is found that while the exact solutions can account for the AC Josephson effect when the coupling strength is significantly less than the voltage, they fail to produce the DC Josephson effect in any practical scenario. To address this fundamental discrepancy, we derive the coupled Ginzburg-Landau (GL) equations for two interconnected superconductors based on BCS theory. These equations reveal that the nonlinear coupling, which is overlooked in the Feynman method, is crucial in describing the spontaneous symmetry breaking in superconductors, a critical factor for achieving the DC Josephson effect. When the coupled GL equations are applied to a double junction, a sawtooth current pattern emerges, a result unattainable via the Feynman approach.
\end{abstract}

\maketitle

\section{Introduction}
For the quantum tunneling between two superconductors (SCs) separated by a thin insulating barrier, Josephson predicts~\cite{josephson1962possible} a current flows through this insulator, known as the Josephson effect. The effect can be categorized into alternating current (AC) and direct current (DC) effects, based on the presence or absence of an external voltage $V$, respectively. These theoretical predictions were swiftly verified by subsequent experiments~\cite{Anderson1963, Rowell1963, Shapiro1963}.

After the Josephson effect was experimentally discovered, various simplified approaches were proposed to understand this effect~\cite{Ferrellmodel, Feynmanbook, Ohtamodel}. Among them, Feynman proposed a renowned approach, which relies on two linearly coupled equations of the `amplitude to find an electron in one side' $\psi_\alpha(t)=\sqrt{\rho_{\alpha}(t)} \exp[i\phi_{\alpha} (t)],(\alpha=L, R)$~\cite{Feynmanbook}, hereafter referred to as the Feynman wave function. Feynman further assumes a constant amplitude (an Ansatz) for $\psi_\alpha(t)$ such that $\rho_R(t)=\rho_L(t) \equiv \rho_0$, which straightforwardly yields the AC Josephson effect. Furthermore, it is claimed by Feynman that the DC Josephson effect is automatically given when the voltage is taken as zero. This oversimplified approach provides an intuitive understanding of the Josephson effect and has been widely adopted in many physics textbooks~\cite{Bassani2005EncyclopediaOC, Superconductivity, Roberto}.

However, Ohta realized that the Ansatz about a constant amplitude of $\psi_\alpha(t)$ conflicts with the conservation of the particle number and fails to meet the linear equations in Feynman approach~\cite {Ohtamodel}. It has been recognized by us that the DC effect is unattainable by the Feynman approach. Despite this, it remains necessary to clarify the specific conditions under which the Ansatz is valid and to further resolve the paradox inherent in the Feynman approach.

In this paper, we start from the exact solution of the Feynman equation, which is easily obtained as that for solving the model of a two-level system. This solution shows that (i) in the weak coupling regime (the coupling strength is smaller than the voltage $K \ll V$), the AC Josephson effect is recovered, which is the same results obtained by the Ansatz in the Feynman approach; (ii) 
neither the DC nor AC effect can be described in the strong coupling regime ($V \ll K$).

Thus, there exists a paradox that the Ansatz in the Feynman approach appears to yield both the AC and the DC Josephson effect without any restrictions, but the exact solution does not. That is why we refer to it as the `Feynman paradox' on the Josephson effect.

To examine the Feynman approach and understand why it fails to give the DC Josephson effect, we derive the coupled Ginzburg-Landau (GL) equation of two SCs from the BCS theory in the mean-field approximation, where the superconducting order parameter is considered as the Feynman wave function. It is discovered that in the coupled GL equation, the nonlinear coupling of the order parameters is included, and such a nonlinear coupling, indicative of the spontaneous symmetry breaking, is crucial for the emergence of the DC Josephson effect. But it was just ignored in the Feynman approach.

The coupled GL equation of order parameter has been used to study the flow stability and superconducting phase transitions~\cite{GL,gor1959microscopic, Coullet1985LargeSI, coullet1987topological, GLS1}. Here, we provide another method to analyze the dynamics of coupled order parameters and it is naturally applied to multiple coupled superconductors and other ultracold atoms systems~\cite{smerzi1997quantum, radzihovsky2010relation, piselli2020josephson, levy2007ac}. As an illustration, we use the coupled GL equation to the double junction, and the DC sawtooth current~\cite{DeLuca, ouassou2017spin, barash2018proximity, tinkham2004introduction, banszerus2024voltage, bozkurt2023double} and richer AC effects are rediscovered. However, the Feynman approach is still powerless to deal with such problems.

This article is structured as follows: Sec.~\ref{sec2} delves into the limitations of the Feynman approach by meticulously solving the Feynman equation. It is clarified that only under weak coupling, i.e. high voltage conditions, the Feynman equation can depict the AC Josephson effect. Additionally, it should be emphasized that the Feynman approach is inadequate for describing the DC effect. Moreover, we point out that the Ansatz used by Feynman is not the strict solution of the equation, and it approximately holds only in the weak coupling limitation. 

In Sec.~\ref{sec3}, we derive the coupled equations of the order parameter of two SCs from the microscopic theory. Starting from a microscopic equation, we show that the Feynman equation is recovered in the weak coupling and reveal the limitations of the Feynman approach. Further in Sec.~\ref{sec4}, we obtain the coupled GL equation of two coupled SCs and find that the nonlinear coupling terms in the coupled GL equation are significant for the DC Josephson effect. Moreover, in Sec.~\ref{sec5}, we extend the coupled GL equation to the double junction and microscopically obtain the AC and DC effects (sawtooth current) in this system.

\section{Feynman paradox to the Josephson effect\label{sec2}}

In this section, we revisit the Feynman approach to the Josephson effect and point out the possible paradox existing here. We solve the Feynman equation exactly. And the exact solutions show that the Feynman approach only yields the AC effect under weak coupling, but does not produce the DC effect at all. 

Feynman proposed an elegant approach to describe the Josephson effect, which includes two linear coupled equations (Feynman equation)~\cite{Feynmanbook}
\begin{equation}
    \begin{aligned}
        \left\{
            \begin{array}{rcl}
            i \frac{\partial}{\partial t} \psi_L(t) &= V_L \psi_L(t) + K \psi_R(t), \\
            i \frac{\partial}{\partial t} \psi_R(t) &= V_R \psi_R(t) + K \psi_L(t), 
            \end{array} \right.
    \end{aligned}
    \label{equation of Josephson effect}
\end{equation}
where $K$ is the phenomenological coupling strength between the two SCs, $V_{\alpha}$ with $\alpha=L, R$ can be regarded as the chemical potential of each SC, $\psi_\alpha$ is the `amplitude to find an electron in one side' (Feynman wave function) and we take $\hbar = 1$ for convenience. The potential difference between SCs is  $V_L-V_R=qV$, where $q =2 e$ is the charge of Cooper pairs. 

It follows from a Laplace transform of Feynman wave function $\psi_{\alpha}(t)$ that the exact solution for the above Feynman equation \eqref{equation of Josephson effect} is
\begin{equation}
\begin{aligned}
    \psi_L(t) &= \left[ (c_+ e^{- i \Omega t} - c_- ) \psi_L(0) - \tilde{K} (1 - e^{- i \Omega t}) \psi_R(0) \right] e^{- i \lambda t},\\
    \psi_R(t) &= \left[(c_+  - c_- e^{- i \Omega t}) \psi_R(0) - \tilde{K} (1 - e^{- i \Omega t}) \psi_L(0)\right] e^{- i \lambda t},
\end{aligned}
\label{ex-solutions}
\end{equation}
where $c_{\pm} = \left( q V \pm \Omega \right)/(2 \Omega)$, $\;\tilde{K} = K/\Omega$ and $\lambda = [(V_L + V_R) - \Omega]/2$ with $\Omega \equiv \sqrt{q^2 V^2 + 4 K^2}$. And $\psi_\alpha(0)$ is the initial value of $\psi_{\alpha}(t)$. Then we rewrite $\psi_{\alpha}(t)$ in terms of amplitude and phase: $\psi_\alpha(t)=\sqrt{\rho_{\alpha}(t)} \exp[i\phi_{\alpha} (t)]$, where the amplitudes are
\begin{equation}
\rho_L(t) = \rho_0+ g(t),\quad \rho_R(t) = \rho_0 - g(t)
\label{rhot}
\end{equation}
with
\begin{equation}
g(t) = \frac{2 K \rho_0}{\Omega^2} \left[\Omega \sin{\phi} \sin{(\Omega t)} + 2 q V \cos{\phi} \sin^2 \left(\frac{\Omega t} {2}\right)\right].
\end{equation}
Here, the initial value of $\psi_{\alpha}(t)$ has been taken as $\psi_\alpha(0) = \sqrt{\rho_0}\exp(i \phi_\alpha)$, and  the phase difference is  $\phi:=\phi_R-\phi_L$.

The current flowing from the left SC to the right SC is defined as ~\cite{Feynmanbook}
\begin{equation}
\begin{aligned}
I(t) &:=\dot{\rho}_L(t) =2K \sqrt{\rho_L(t)\rho_R(t)} \cdot \sin[\phi_R(t)-\phi_L(t)]\\
&= 2 K \rho_0 \sqrt{\frac{q^2 V^2}{\Omega^2} \cos^2{\phi} + \sin^2{\phi}} \cdot \sin(\Omega t + \theta)
\label{equation of current}
\end{aligned}
\end{equation} 
with $\theta = \arctan[\Omega \tan{\phi}/(qV)]$. It is worth noting that the current Eq. \eqref{equation of current} can not directly result in the AC or DC Josephson effect. Therefore, below we will discuss the effect of current under different coupling strengths.
\subsection{Weak coupling}
In the weak coupling regime, where the coupling strength is less than the applied voltage  $K\ll V$,  $\rho_{\alpha}(t)$ is approximated as $\rho_L(t) \approx \rho_R(t) =\rho_{0}$ [see Eq. \eqref{rhot}]. And the current \eqref{equation of current} also degenerates to give the AC Josephson effect
\begin{equation}
    \begin{aligned}
        I(t) \approx 2 K \rho_{0} \sin{(q V t + \phi)}.
    \end{aligned}\label{equation of current in Feynman's book}
\end{equation}
This is just the main result given by the Feynman approach [see Appendix~\ref{sec: appendixB}]. In this sense, Feynman's result is only an approximation of the exact solution in weak coupling $K\ll V$, and it is observed directly from a perturbation method as the order of $K/V$ [see Appendix~\ref{sec: appendixC}]. Consequently, the AC effect based on \eqref{equation of current in Feynman's book} is accurately predicted by the Feynman approach. 

However, the Feynman approach cannot describe the DC Josephson effect because the zero voltage is undoubtedly much beyond the weak coupling regime ($K \ll V$). This is the crux of Feynman's paradox.

\subsection{Strong coupling}
In the strong coupling regime ($K\gg V $), the higher order terms of $(V/K)$ as a small quantity are neglected. The current \eqref{equation of current} is simplified as
\begin{equation}
    \begin{aligned}
        I(t) \approx 2 K \rho_0 \sin \phi \sin(2 K t + \theta)
    \end{aligned}
    \label{strong-coupling}
\end{equation}
which oscillates over time but the oscillation frequency of the current is twice the coupling strength. Obviously, this has seriously deviated from the AC Josephson effect. Even if the voltage is zero $V=0$, the current still maintains oscillation 
\begin{equation}
I(t) = 2 K \rho_0 \sin{\phi} \cos{2 K t}.
\label{zero-voltage}    
\end{equation}
It is seen from Eqs. (\ref{strong-coupling}, \ref{zero-voltage}) that the Feynman equation does not describe the AC and DC effect when the voltage is small.

Therefore, the Feynman equation only depicts the AC effect in the weak coupling limit $K\ll V$. Therefore, below we will start from microscopic theory to study why Feynman equation \eqref{equation of Josephson effect} cannot work well in predicting the DC Josephson effect in Sec.~\ref{sec3}

\section{The Feynman approach from BCS Theory\label{sec3}}
In this section, we show that the Feynman equation can be derived from the BCS theory only in the weak coupling conditions where the Feynman wave function is regarded as the superconducting order parameter. In our approach, the phenomenological coefficient of linear coupling in the Feynman equation is directly determined by the microscopic parameters in the BCS theory.

\subsection{Feynman wave function with BCS}
\begin{figure}
    \centering
    \includegraphics[width=8.5cm]{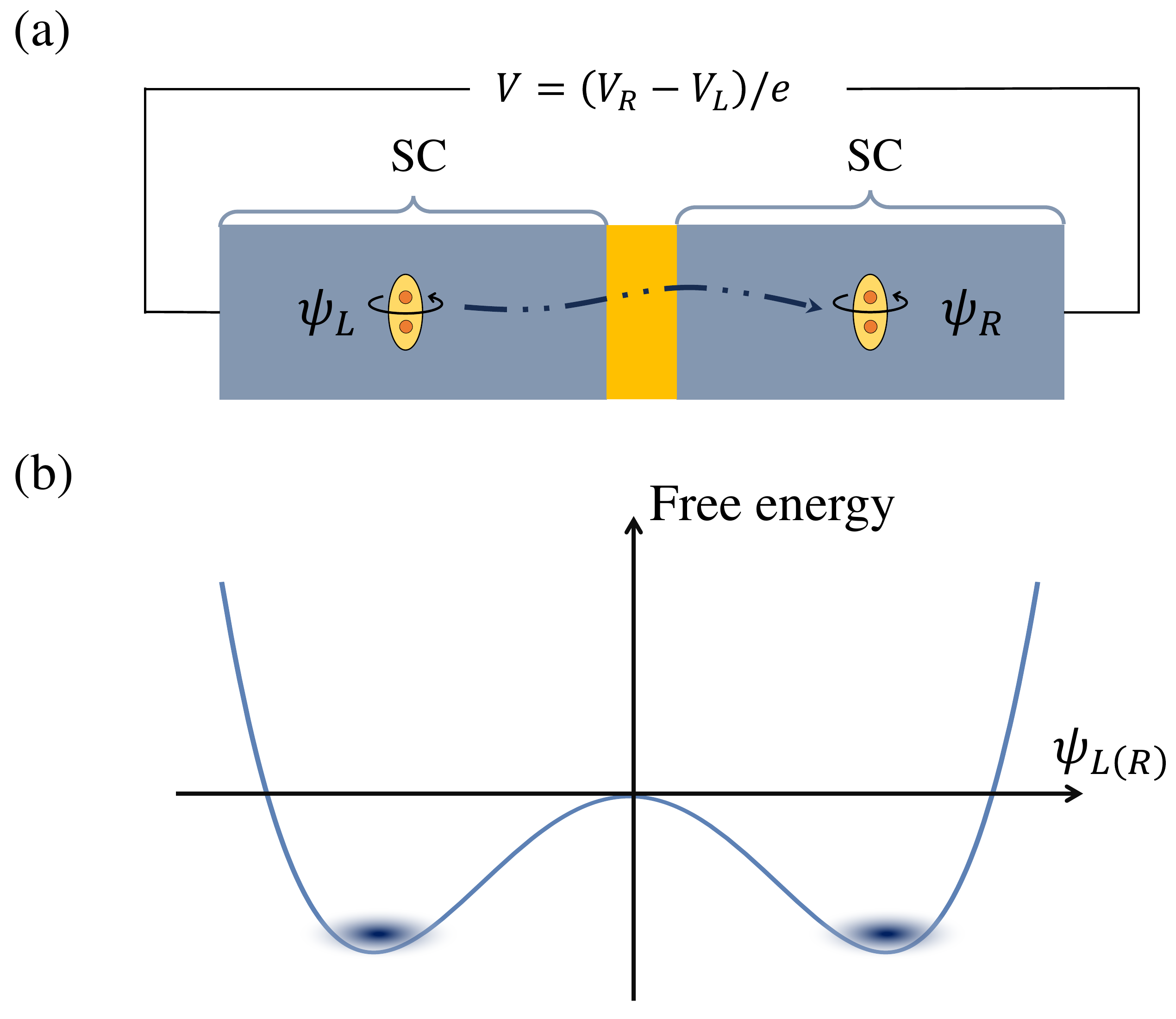}
    \caption{\label{Josephson junction} (a) The two SCs are linked by an insulating barrier in the applied voltage $V$ and the interaction between them is via Cooper pairs tunneling. Such coupled SCs are characterized by the coupled Ginzburg-Landau equation. (b) The free energy of the spontaneous symmetry breaking.}
\end{figure}

The Hamiltonian of the $s$-wave SC is read as
\begin{equation}
H_{\alpha} = \sum_{\textbf{k},\sigma} \varepsilon_{\alpha,\textbf{k}} \hat{c}_{\alpha,\textbf{k} \sigma}^{\dagger} \hat{c}_{\alpha,\textbf{k} \sigma}- U \sum_{\textbf{k,q}} \hat{c}_{\alpha,\textbf{k} \uparrow}^{\dagger} \hat{c}_{\alpha,-\textbf{k}\downarrow}^{\dagger} \hat{c}_{\alpha,-\textbf{q}\downarrow} \hat{c}_{\alpha, \textbf{q}\uparrow}.
\label{Hamiltonian of BCS}
\end{equation}
Here, $\varepsilon_{\alpha,\textbf{k}} = \textbf{k}^2/2m - \epsilon_F- V_{\alpha}$ is the kinetic energy of the electron measured from the Fermi level $\epsilon_F$, where $V_{\alpha}$ is the chemical potential of the each SC introduced by the applied voltage. And $U$ is the effective attractive interaction between electrons induced by the exchange of phonons. Notice that we have considered that the effective attractive interactions are identical for convenience. Two SCs are linked by an insulator, and the interaction between them is via the tunneling of the Cooper pair~\cite{Wallace1965, lee1971theory} as shown in Fig. \ref{Josephson junction}(a)
\begin{equation}
H_T= - g \sum_{\textbf{k,q}} \left(\hat{c}_{L,\textbf{k} \uparrow}^{\dagger} \hat{c}_{L,-\textbf{k} \downarrow}^{\dagger} \hat{c}_{R,-\textbf{q} \downarrow} \hat{c}_{R,\textbf{q} \uparrow} + \rm{h.c.}\right).
\label{Hamiltonian of BCS}
\end{equation}
Here, $g$ is the tunneling strength between SCs.

The current flowing into the left SC is defined by a change of the particle number as
\begin{equation}
I(t) = 4 e g \sum_{\textbf{k,q}} \Im \left[\bra{ \rm{BSC} }\hat{c}_{L,\textbf{k}\uparrow}^{\dagger}  \hat{c}_{L,-\textbf{k} \downarrow}^{\dagger} \hat{c}_{R, -\textbf{q} \downarrow} \hat{c}_{R,\textbf{q} \uparrow} \ket{ \rm{BSC} } \right],
\label{current equation}
\end{equation}
where $e$ is the elementary charge. The initial state of the system is taken as the product of the BCS ground states of two SCs at zero temperature~\cite{Wallace1965, lee1971theory}
\begin{equation}
\begin{aligned}
\ket{\rm{BCS}} \equiv \ket{\rm{BCS}(\phi_L)} \otimes \ket{\rm{BCS}(\phi_R)}, 
\end{aligned}
\end{equation}
where 
\begin{equation}
\ket{\rm{BCS}(\phi_{\alpha})}:= \prod_k \left(u_{\alpha, k} + v_{\alpha, k} e^{i \phi_{\alpha}} \hat{c}_{\alpha,\textbf{k} \uparrow}^{\dagger} \hat{c}_{\alpha,- \textbf{k} \downarrow}^{\dagger} \right) \ket{\rm{vac}}. \label{BCS ground state}
\end{equation}
The order parameter of each SC is defined as
\begin{equation}
\begin{aligned}
\psi_{\alpha}(t) = \sum_k \langle \hat{c}_{\alpha,-\textbf{k} \downarrow} \hat{c}_{\alpha,\textbf{k} \uparrow} \rangle \equiv \sqrt{\rho_{\alpha}(t)} \exp[i \phi_{\alpha}(t)].
\end{aligned}
\end{equation} 
In the mean-field approximation, the current Eq. \eqref{current equation} is formally approximated as 
\begin{equation}
\begin{aligned}
I(t) &\approx 4 e g \Im[\psi_L^*(t) \psi_R(t)] = 4 g \sqrt{\rho_L(t) \rho_R(t)} \sin [\delta(t)],
\end{aligned}\label{current from the BCS}
\end{equation}
where $\delta(t) = \phi_R(t) - \phi_L(t)$ is the phase difference between the two SCs at time $t$. It is worth noting that the current defined in the Feynman approach [see Eq. \eqref{equation of current}] is in coincidence with this current form \eqref{current from the BCS}. In this sense, we regard the order parameter of each SC as the Feynman wave function.

\subsection{Feynman equation from BCS}
Explicitly, the current is determined by the phase and amplitude of the order parameter. To calculate $\rho_{\alpha}(t)$ and $\phi_{\alpha}(t)$, we consider the Heisenberg equation of $\psi_\alpha$. In the mean-field approximation, and focusing solely on the electrons near the Fermi surface on conduction, i.e., $\textbf{k}^2/2m \approx \epsilon_F$, the equations of $\psi_{\alpha}$ become
\begin{equation}
\begin{aligned}
i \dot{\psi}_L &\approx - 2 V_L \psi_L - U \left[N_L(t) - N \right] \psi_L - g \left[N_L(t) - N \right] \psi_R,\\
i \dot{\psi}_R &\approx - 2 V_R \psi_L - U \left[N_R(t) - N \right] \psi_R - g \left[N_R(t) - N \right] \psi_L,
\end{aligned}\label{eq of delta}
\end{equation}
where $N_{\alpha}(t) \equiv \langle \hat{N}_{\alpha}(t) \rangle$ is the average number of electrons in SC, and $N_{\alpha}(0)=\langle \hat{N}_{\alpha}(0) \rangle \equiv N + p_{\alpha}$ with $p_{\alpha} = \sum_k (v_{\alpha,k}^2 - u_{\alpha,k}^2)$ and $N = \sum_k 1$. 

In the case of weak coupling ($g \ll V$), $N_{\alpha}(t)$ can be expanded to the first order of $g/V$ as: 
\begin{equation}
\begin{aligned}
N_L(t) \approx N + p_{L} - \frac{g}{V} f_{L}(t),\\
N_R(t) \approx N + p_{R} + \frac{g}{V} f_{R}(t),
\end{aligned} \label{expand of Jz}
\end{equation}
where $f_{\alpha}(t):= \left( \sum_k u_{\alpha,k} v_{\alpha,k} \right)^2 \left[ \cos{(2 V t + \phi)} - \cos{\phi} \right]$. When the coupling strength is so much less than the voltage that the first order of $g/V$ is neglected, it is followed from  \eqref{eq of delta} that
\begin{equation}
\begin{aligned}
i \dot{\psi}_L &\approx - 2 (V_L + U p_{L}) \psi_L  - g p_L \psi_R,\\
i \dot{\psi}_R &\approx - 2 (V_R + U p_R) \psi_R  - g p_R \psi_L,
\end{aligned}\label{Feynman equation form BCS}
\end{equation}
which is exactly the Feynman equation if $p_L = p_R \equiv p$ and the phenomenological parameter is determined as $K = - g p $. It is found that the condition of the acquirement of Eq. \eqref{Feynman equation form BCS} $g \ll V$ is consistent with the condition for the Feynman approach to obtain the AC Josephson effect. Consequently, without an applied voltage, the Feynman equation is not derived from BCS theory through perturbation theory, not to mention obtaining the DC effect through the Feynman equation. 
Therefore, we then attempt to modify the Feynman approach to give the DC effect.

\section{Nonlinearity for the Josephson effect\label{sec4}}
In this section, we present the coupled GL equation of the two SCs to describe the Josephson effect. Furthermore, we analyze the role of spontaneous symmetry breaking, as indicated by nonlinearity in the equations, in realizing the Josephson effect.

Notice that in the mean field approximation and the conditions of the perturbation expansion: $g \ll \Delta_{\alpha}$ where $\Delta_\alpha:= \sum_{\textbf{k}}|\bra{ \rm{BCS}} \hat{c}_{\alpha,-\textbf{k} \downarrow}(0) \hat{c}_{\alpha,\textbf{k}\uparrow}(0) \ket{\rm{BCS}} \rangle|=\sum_{k} u_{\alpha,k} v_{\alpha,k}$ is the superconducting energy gap for SC-$\alpha$, the relationship between $N_{\alpha}$ and  $\abs{\psi_{\alpha}(t)}^2$ is
\begin{equation}
\abs{\psi_{\alpha}(t)}^2 \approx \sum_{\textbf{k,q}} \langle \hat{c}_{{\alpha},\textbf{k} \uparrow}^{\dagger} \hat{c}_{{\alpha},-\textbf{k} \downarrow}^{\dagger} \hat{c}_{{\alpha}, -\textbf{q} \downarrow} \hat{c}_{{\alpha},\textbf{q} \uparrow} \rangle \approx N_{\alpha}(t) - N + \Delta_{\alpha}^2\label{Jz and J+J-}
\end{equation}
It follows from Eq. (\ref{eq of delta}, \ref{Jz and J+J-}) that the coupled GL equation of the order parameter is obtained as 
\begin{equation}
\begin{aligned}
i \dot{\psi}_L &= 2 \mu_L \psi_L - U |\psi_L|^2 \psi_L + G_L \psi_R - g |\psi_L|^2 \psi_R,\\
i \dot{\psi}_R &= 2 \mu_R \psi_R - U |\psi_R|^2 \psi_R + G_R \psi_L - g |\psi_R|^2 \psi_L
\end{aligned} \label{coupled GL equation}
\end{equation}
where $\mu_{\alpha}:= U \Delta_{\alpha}^2/2 - V_{\alpha}$. In addition to the linear coupling between two order parameters, represented by $G_{\alpha} = g \Delta_{\alpha}^2$ in the above equation, there exists the nonlinear coupling $U$ and $g$. This nonlinear term characterizes the condensation phenomenon in SCs, and it also encapsulates the spontaneous symmetry breaking within SCs, as depicted in Fig. \ref{Josephson junction}(b). It is directly seen that a solution of Eq. \eqref{coupled GL equation} is 
\begin{equation}
\begin{aligned}
\psi_L(t) = \Delta_L e^{i(2 V_L t + \phi_L)},\;\;
\psi_R(t) = \Delta_R e^{i(2 V_R t + \phi_R)}.
\end{aligned}
\end{equation}
According to the solution, the current is obtained by \eqref{equation of current} as
\begin{equation}
\begin{aligned}
I(t) = I_0 \sin{\left( 2eVt + \phi \right)},
\end{aligned}\label{Josephson effect}
\end{equation}
where $V = (V_R - V_L)/e$ is the applied voltage difference, $\;I_0 = 4 e g \Delta_L \Delta_R$ is the critical current of Josephson junction and $\phi=\phi_R-\phi_L$ is the initial phase difference of the two SCs. The current \eqref{Josephson effect} precisely describes the AC Josephson effect. Simultaneously, the DC effect manifests when the applied voltage difference is zero $V=0$. Therefore, the coupled GL equation is capable of describing both the DC and AC Josephson effects. 

\begin{figure}
    \centering
    \includegraphics[width=8.8cm]{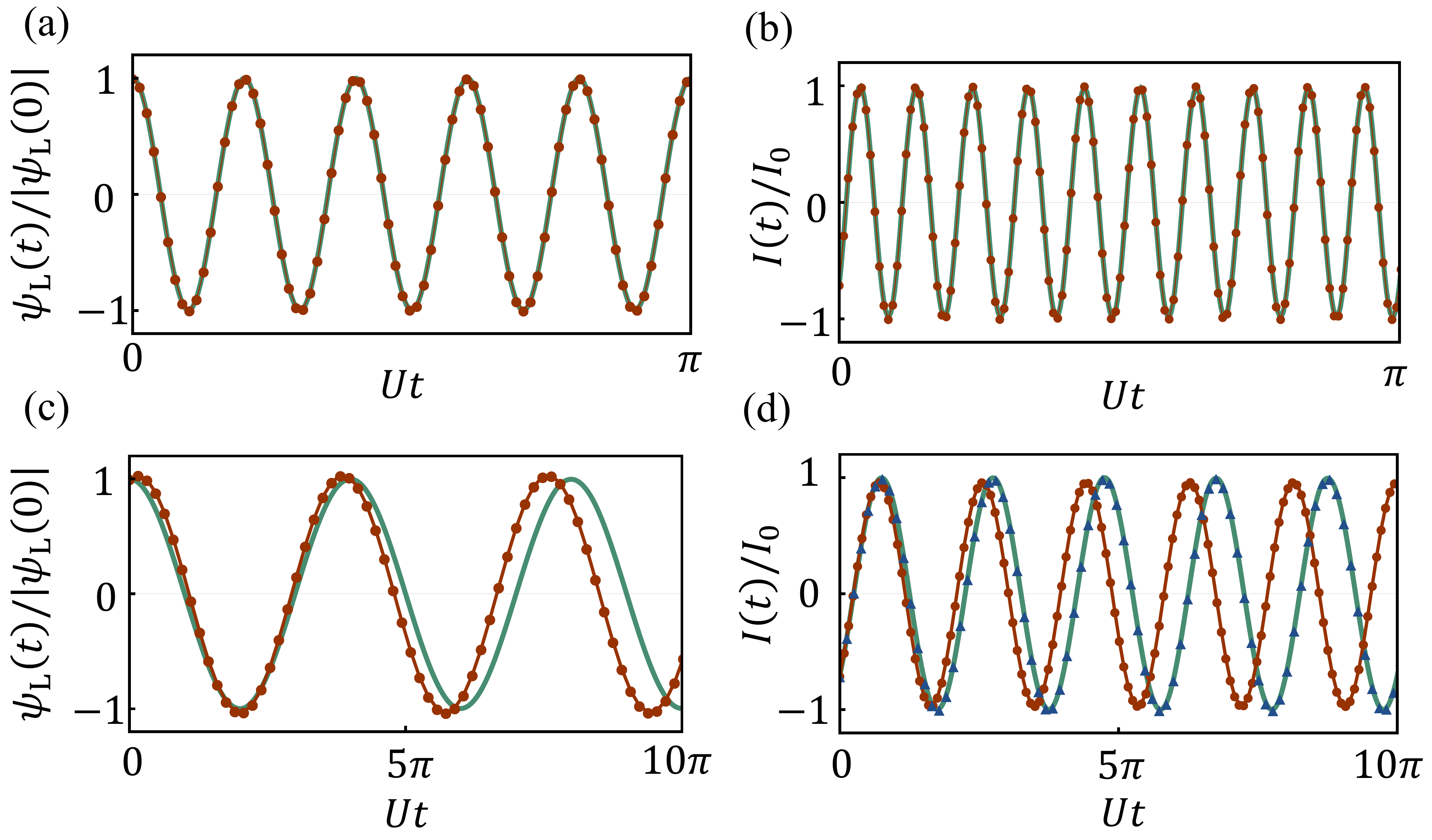}
    \caption{\label{linearandnonw} The time evolution of order parameters $\psi_{\alpha}(t)$ and current $I(t)$ for the given phase difference $\phi = \pi/4$ in the different strength of coupling scheme: $V/U = 10$ in (a, b) and $V/U = 0.5$ in (c, d) with $g/U= K/U = 0.2$. The numerical result of order parameter and the current is obtained by coupled GL equation \eqref{coupled GL equation} [green solid line] and the Feynman equation \eqref{equation of Josephson effect} [red-filled-circle line], where $\abs{\psi_L(0)}$ and $I_0$ represent the amplitude of the initial value of $\psi_L(t)$ and the current. Besides, the dotted line composed by the blue-filled triangle in (d) reflects the AC Josephson current.}
\end{figure}

Moreover, we compare the numerical result between the coupled GL equation \eqref{coupled GL equation} and the Feynman equation \eqref{equation of Josephson effect}. It is discovered that the former can better describe the Josephson effect than the Feynman equation as shown in Fig. \ref{linearandnonw}. There is no difference between the result of the coupled GL equation (solid line) and the Feynman equation (red-filled-circle line) under the weak coupling ($g/V = K/V = 0.02$) [in Fig. \ref{linearandnonw} (a) and (b)]. But in the stronger coupling limits ($g/V = K/V = 0.4$), Fig. \ref{linearandnonw}(c) and (d) show that the coupled GL equation provides a more practical description of the Josephson effect than the Feynman equation due to the good fitness between the coupled GL equation [green solid line in Fig. \ref{linearandnonw}(d)] and the Josephson current $\sin{(2eVt-\phi)}$ [blue-filled-triangle dotted line in Fig. \ref{linearandnonw}(d)].

As a result, the coupled GL equation \eqref{coupled GL equation} of the coupled SCs offers a more realistic description of the Josephson effects even in the stronger coupling limit. However, the Feynman equation ignored the nonlinear coupling in the coupled GL equation which is essential for reaching the DC Josephson effect.

\section{Sawtooth current in double junction\label{sec5}}

\begin{figure}
    \centering
    \includegraphics[width=8.6cm]{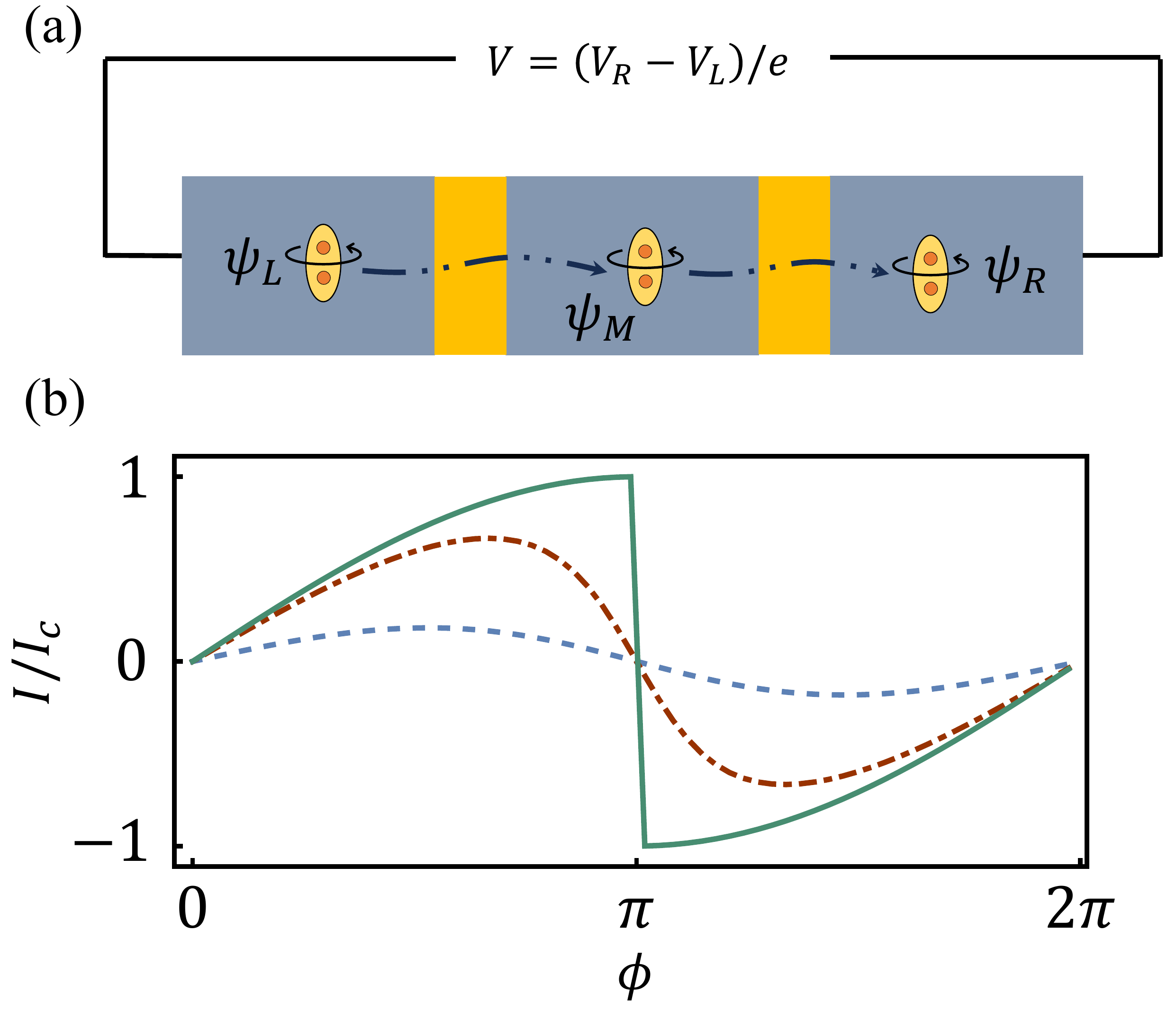}
    \caption{\label{double junction} (a). The schematic diagram of a double junction is made up of three SCs linked by two insulating barriers in the applied voltage $V$ and the interaction between them is via Cooper pairs tunneling. (b). Current through a double junction at various relative coupling energies: $\rho = 0.1$ (blue dashed line), $\rho = 0.5$ (red-dot-dashed line), and $\rho = 1$ (green solid line).}
\end{figure}

In the above section, it has been shown that the Feynman approach cannot describe the DC effect in the Josephson junction, let alone the multiple junctions. However, the coupled GL equation can be used generally to study the current effects of multiple coupled SCs. In this section, we apply the coupled GL equation to the three coupled SCs (double junction), as shown in Fig. \ref{double junction}. Unlike the usual Josephson current, the sawtooth current~\cite{DeLuca, ouassou2017spin, barash2018proximity, tinkham2004introduction, bozkurt2023double} appears as the DC effect, and relevant experiments are also reported recently~\cite{banszerus2024voltage}. This fully reflects that the nonlinear coupling in our equation signifies the spontaneous symmetry breaking which is essential for the emergence of the DC effect in the double junction.

The double junction is described by $H= H_L + H_M + H_R + H_T$, where the three different $s$-wave SCs are denoted $\alpha = L, M, R$ as 
\begin{equation}
H_{\alpha} = \sum_{\textbf{k},\sigma} \varepsilon_{\alpha,\textbf{k}} \hat{c}_{\alpha,\textbf{k} \sigma}^{\dagger} \hat{c}_{\alpha,\textbf{k} \sigma}- U \sum_{\textbf{k,q}} \hat{c}_{\alpha,\textbf{k} \uparrow}^{\dagger} \hat{c}_{\alpha,-\textbf{k}\downarrow}^{\dagger} \hat{c}_{\alpha,-\textbf{q}\downarrow} \hat{c}_{\alpha, \textbf{q}\uparrow},\\
\end{equation}
Similarly, $\varepsilon_{\alpha,\textbf{k}} := \textbf{k}^2/2m - \epsilon_F- V_{\alpha}$ represents the kinetic energy of the electron measured from the Fermi level $\epsilon_F$, and $V_{\alpha}$ is the chemical potential of each SC. And the interaction between SCs on both sides and the middle SC still through the tunneling of Copper pairs
\begin{equation}
H_T= - \sum_{\lambda = L,R} \sum_{\textbf{k,q}} g_{\lambda} c_{\lambda,\textbf{k} \uparrow}^{\dagger} c_{\lambda,-\textbf{k} \downarrow}^{\dagger} c_{M,-\textbf{q} \downarrow} c_{M,\textbf{q} \uparrow} + \rm{h.c.}.
\end{equation}

In the initial state that the product of the BCS ground states of three SCs
\begin{equation}
\begin{aligned}
\ket{\rm{BCS}} \equiv \ket{\rm{BCS}(\phi_L)} \otimes \ket{\rm{BCS}(\phi_M)} \otimes \ket{\rm{BCS}(\phi_R)},
\end{aligned}\label{state of three SCs}
\end{equation} 
like Eq. \eqref{current from the BCS}, the current  flowing from middle into left SC is 
\begin{equation}
I(t) \approx 4 e g_L \Im[\psi_L^*(t) \psi_R(t)]. \label{current of three Scs}
\end{equation}
To obtain the above current, we extend the coupled GL to three coupled SCs, and the order parameter $\psi_{\alpha}(t)$ satisfies 
\begin{equation}
\begin{aligned}
i \dot{\psi}_L &= 2 \mu_L \psi_L - U \abs{\psi_L}^2 \psi_L + G_{LL} \psi_M - g_L \abs{\psi_L}^2 \psi_M,\\
i \dot{\psi}_M &= 2 \mu_M \psi_M - U \abs{\psi_M}^2 \psi_M + \sum_{\alpha = L,R} \left( G_{\alpha M} - g_{\alpha} \abs{\psi_M} \right) \psi_{\alpha},\\
i \dot{\psi}_R &= 2 \mu_R \psi_R - U \abs{\psi_R}^2 \psi_R + G_{RR} \psi_M - g_L \abs{\psi_R}^2 \psi_M
\end{aligned} \label{equation of three scs}
\end{equation}
where $\mu_{\alpha}:= U \Delta_{\alpha}^2/2 - V_{\alpha}$ with the superconducting gap $\Delta_{\alpha}=\sum_{k} u_{\alpha,k} v_{\alpha,k}$.
And the linear coupling $G_{\alpha \beta} = g_{\alpha} \Delta_{\beta}$ and the nonlinear coupling $U, g_{\alpha}$ are still included. Similarly, there exists a solution of Eq. \eqref{equation of three scs} as
\begin{equation}
\begin{aligned}
\psi_{\alpha}(t) = \Delta_{\alpha} e^{i(2 V_{\alpha} t + \phi_{\alpha})}
\end{aligned}\label{solution of three SCs}
\end{equation}
where $\phi_{\alpha}$ denotes the initial phase for each SC. By redefining the phase and potential differences as $\tilde{\phi}_{L(R)}\equiv \phi_{L(R)} - \phi_M  $ and $ \tilde{V}_{L(R)}\equiv V_{L(R)} - V_M$ respectively. Then the current \eqref{current of three Scs} is reduced to 
\begin{equation}
I(t) = - I_{0,L} \sin(2 \tilde{V}_L t + \tilde{\phi}_L).
\end{equation}

In the case of no external voltage ($\tilde{V}_\alpha= 0$), it is seen from [$\tilde{\phi}_{\alpha}(t) = 2 \tilde{V}_{\alpha} t + \tilde{\phi}_{\alpha}$] that the phase difference of the left and right SC must be fixed (i.e. $\tilde{\phi}_R - \tilde{\phi}_L = \phi$), which can be adjusted in experiment~\cite{banszerus2024voltage, gupta2023gate}. At this time, the current becomes 
\begin{equation}
I =I_{0,L} \sin (\phi - \tilde{\phi}_R).
\end{equation}

The energy of the system in the initial state is 
\begin{equation}
E = E_0 - E_{c,L} \cos(\tilde{\phi}_R - \phi) - E_{c,R} \cos{\tilde{\phi}_R},
\end{equation}where $E_0$ is independent of the phase of SC and $E_{c,\alpha} = 2 g_{\alpha} \Delta_{\alpha} \Delta_M$. It is observed that the product of the three BCS states \eqref{state of three SCs} is not the ground state of the double junction, but it can be regarded as a variational ground state. Obviously, besides the observed phase difference $\phi$, the energy also depends on another redundant variable $\tilde{\phi}_R$. Therefore, we need to further minimize the energy with respect to $\tilde{\phi}_R$ so that the optimized state is much closer to the actual ground state of the double junction. By minimizing the energy with respect to $\tilde{\phi}_R$, the phase relationship is obtained as  
\begin{equation}
\frac{\partial E}{\partial \tilde{\phi}_R} = 0 \Rightarrow \tilde{\phi}_R = \frac{\phi}{2} - \arctan(\frac{1 - \rho}{1 + \rho} \tan \frac{\phi}{2}) + n \pi.
\label{extreme value conditions}
\end{equation}
where $\rho = E_{c,L}/E_{c,R}$ is the relative coupling energy. And the second derivative  of the energy at this extreme point is 
\begin{equation}
\frac{\partial^2 E}{\partial \tilde{\phi}_R^2} = (-1)^n \rm{sign} \left(\cos \frac{\phi}{2}\right)g(\phi)   \;\;(n \in \mathbb{N}) \label{extreme value conditions b}
\end{equation}
where
\begin{equation}
g(\phi) = (E_{c,L} + E_{c,R}) \sqrt{\cos^2({\phi}/{2}) + \frac{(1-\rho)^2}{(1+\rho)^2} \sin^2 ({\phi}/{2})}
\end{equation}
is constantly greater than zero. Therefore, the condition of minimizing the energy at $\tilde{\phi}_R$, that is $\partial^2 E/\partial \tilde{\phi}_R^2 >0$ seen from Eq. \eqref{extreme value conditions b} requires 
\begin{equation}
(-1)^n  = \rm{sign} \left(\cos \frac{\phi}{2}\right). \label{minimal}
\end{equation}

The above extreme value conditions (\ref{extreme value conditions}, \ref{minimal}) are equivalent to the optimization of energy on average superconducting phase $(\tilde{\phi}_L + \tilde{\phi}_R)/2$~\cite{tinkham2004introduction, DeLuca}. Under these extreme value conditions, the current of the system becomes
\begin{equation}
I=I_c \frac{2 \rho}{(1 + \rho)^2} \frac{\sin \phi}{\sqrt{1 - \frac{4 \rho}{(1 + \rho)^2} \sin^2 \frac{\phi}{2}}}
\end{equation}
with $I_c = 2 e \Delta_M (\Delta_L g_L + \Delta_R g_R)$, and the current curve with the phase is shown in Fig. \ref{double junction} (b). When $g_L \Delta_L = g_R \Delta_R \equiv g \Delta\Rightarrow\rho = 1$, the current is simplified as
\begin{equation}
I = I_c \rm{sign} \left( \cos \frac{\phi}{2}\right) \sin \frac{\phi}{2}
\end{equation}
which gives the `sawtooth' current (green solid line in Fig. \ref{double junction} (b)) of the double junction~\cite{DeLuca, ouassou2017spin, barash2018proximity, banszerus2024voltage}. As a result, we derive microscopically the sawtooth current in the double junction through the coupled GL equation.

When an external voltage is applied such that $\tilde{V}_R - \tilde{V}_L = e V$, the phase of the left or right SC changes over time as described by $\tilde{\phi}_{L(R)}(t) = 2 \tilde{V}_{L(R)} t + \tilde{\phi}_{L(R)}$. As the initial phase difference is constant $\tilde{\phi}_R - \tilde{\phi}_L = \phi$, the energy in the initial state has been optimized to the minimal with the extreme value conditions Eq. \eqref{extreme value conditions}. By resetting the energy zero point such that $\tilde{V}_L + \tilde{V}_R = 0$, the time-dependent current is obtained as
\begin{equation}
\begin{aligned}
I(t)= I_c \frac{2 \rho}{(1 + \rho)^2} \sqrt{\frac{\rho^2 + 2 \rho \cos \phi + 1}{1 - \frac{4 \rho}{(1 + \rho)^2} \sin^2 \frac{\phi}{2}}} \sin(e V t + \varphi)
\end{aligned}
\end{equation}
with $\varphi = \arctan [\sin \phi/(\rho + \cos \phi)]$, which characterizes the AC effect in the double junction. Setting $V = 0$, the DC effect is obtained.

\section{concluding remarks \label{sec6}}

Feynman's approach to the Josephson effect has offered intuitive insight and has been extensively utilized in numerous textbooks and monographs. However, a paradox arises from the rigorous solutions to the two-component equation in this approach: the AC Josephson effect can only be approximately realized in the weak coupling case, where $K$ is much less than $V$. Consequently, the Feynman approach falls short in achieving the DC effect, as the zero-voltage characteristic inherent to the DC effect does not emerge under the weak coupling condition necessary for the AC effect.

To reconcile this paradox, we have expanded upon the Feynman approach by deriving the coupled Ginzburg-Landau (GL) equations for two interconnected superconductors (SCs). The exact solution of the resulting generalized Feynman equation, which includes a nonlinear GL term, provides an accurate description of both the AC and DC Josephson effects. Our studies underscore the significance of the nonlinear coupling of the order parameter in the coupled GL equations, which represents the spontaneous symmetry breaking of the SC, a crucial factor in achieving the DC Josephson effect. Unfortunately, such nonlinear coupling was overlooked in the original Feynman approach. It is thereby evident that the initial Feynman approach cannot accommodate more complex systems with multiple coupled superconductors to effectively yield the DC effect.

In contrast, the coupled GL equations prove to be functional in such cases, for instance, predicting the AC and DC effects in the system of double junctions. In future investigations, the coupled GL equations will be further employed to explore other effects, such as the nonreciprocal superconducting current (where the forward critical current does not equal the reverse one)~\cite{hu2007proposed, wu2022field, gupta2023gate, souto2022josephson, gutfreund2023direct}, among others.

\section*{acknowledgment \label{sec7}}

The author would like to thank Xin Yue for his valuable advice on the manuscript. This work was supported by the National Natural Science Foundation of China (NSFC) (Grant No. 12088101) and NSAF No.U2330401.

\twocolumngrid
\appendix   
\setcounter{table}{0}   
\setcounter{figure}{0}
\renewcommand{\thetable}{A\arabic{table}}
\renewcommand{\thefigure}{A\arabic{figure}}

\section{Revisit the Feynman approach\label{sec: appendixB}}

The Feynman approach~\cite{Feynmanbook} is summarized as follows. The Feynman equation includes two linear coupled equations 
\begin{equation}
    \begin{aligned}
        \left\{
            \begin{array}{rcl}
            i \frac{\partial}{\partial t} \psi_L(t) &= V_L \psi_L(t) + K \psi_R(t), \\
            i \frac{\partial}{\partial t} \psi_R(t) &= V_R \psi_R(t) + K \psi_L(t), 
            \end{array} \right.
    \end{aligned}
    \label{Feynmaneq}
\end{equation}
in which the $\psi_L(t)$ and $\psi_R(t)$ can be decomposed into the amplitude and phase $\psi_{\alpha} = \sqrt{\rho_{\alpha}(t)} \exp[i \phi_{\alpha}(t)]$. Then the evolution of phase and amplitude are
\begin{equation}
\begin{aligned}
&\dot{\rho}_L(t)=2 K \sqrt{\rho_L(t) \rho_R(t)} \sin \delta(t), \\
&\dot{\rho}_R(t)=-2 K \sqrt{\rho_L(t) \rho_R(t)} \sin \delta(t), \\
&\dot{\phi}_L(t)=-K \sqrt{\frac{\rho_R(t)}{\rho_L(t)}} \cos \delta(t) - \frac{q V}{2},\\
&\dot{\phi}_R(t)=-K \sqrt{\frac{\rho_L(t)}{\rho_R(t)}} \cos \delta(t) + \frac{q V}{2},
\end{aligned} \label{equation of amplitude and phases}
\end{equation}
where $\delta(t) = \phi_L(t) - \phi_R(t)$ is the phase difference. 

The current flowing the left SC in this system is defined as $I(t) \equiv \dot{\rho}_L(t) = 2K \sqrt{\rho_R(t) \rho_L(t)} \sin \delta(t)$. By taking an Ansatz by Feynman as $\rho_L(t) = \rho_R(t) = \rho_0 \equiv \rm{constant} $, the above current is simplified as
\begin{equation}
I = I_0 \sin{\delta(t)}, \label{current form in Feynman}
\end{equation}
where the amplitude of the current is $I_0 \equiv 2 K \rho_0$. With the evolution equations of $\phi_L(t)$ and $\phi_R(t)$ in Eq. \eqref{equation of amplitude and phases}, the equation of the phase difference $\delta(t)$ becomes 
\begin{equation}
\delta(t) = \delta(0) + \int_0^t [\dot{\phi}_R(\tau) - \dot{\phi}_L(\tau)] d\tau = \delta(0) + q V t,
\end{equation}
which is the current phase relation in the Josephson effect. The DC effect can be directly obtained by taking $V = 0$ ~\cite{Feynmanbook}. Therefore, Feynman obtained both DC and AC effects through the Ansatz. 

However, the Ansatz is not the strict solution to Eq. \eqref{equation of amplitude and phases}. It is easy to see that if $\rho_L$ and $\rho_R$ are identical, then $\rho(t) \sin{\delta(t)}=0$ should be valid according to the evolution equation of amplitude in Eq. \eqref{equation of amplitude and phases}, which means $\delta(t) = n \pi\;(n \in \mathbb{N})$ when $\rho(t) \neq 0$, and the current phase relation cannot hold. Therefore, $\rho_L$ and $\rho_R$ should not be identical strictly. In the Appendix~\ref{sec: appendixC}, we will demonstrate from a perturbative perspective that this Ansatz by Feynman is only the zeroth-order outcome~\cite{romeo2022cooper} so that it cannot be used to predict the DC Josephson effect.

\section{Analyzing the Ansatz of Feynman\label{sec: appendixC}}

In this section, we examined the Ansatz in Feynman approach through the perturbation theory and showed that it is only the zeroth order solution of the Feynman equation just as pointed out in ref. \cite{romeo2022cooper}. 

We expand $\rho_{\alpha}(t)$ according to the order of $K$, and its zero-order is
\begin{equation}
\rho_{\alpha}(t) = \rho_{\alpha}^{(0)}, \;\;\alpha = L,R.
\end{equation}
Then the amplitude equations of $\rho_{\alpha}(t)$ become
\begin{equation}
\dot{\rho}_{\alpha}^{(0)}(t) = 0\Rightarrow{\rho_L^{(0)} = \rho_R^{(0)} = \rho_0 },
\label{B2}
\end{equation}
and the equation of phase can be simplified as
\begin{equation}
\dot{\delta}(t) = \dot{\phi}_L(t) - \dot{\phi}_R(t) = - q V.
\label{B3}
\end{equation}
It is seen from Eq. \eqref{B2} and \eqref{B3} that Feynamn's Ansatz is just the lowest-order solution of the Feynman equation. 

Further, we expand $\rho_{\alpha}(t)$ to the order of $K$, that is $\rho_\alpha (t) \approx \rho_0 + K \rho_\alpha^{(1)}(t)$ with which the first equation in Eq. \eqref{equation of amplitude and phases} can be written as
\begin{equation}
\begin{aligned}
\dot{\rho}_L(t) &= \dot{\rho}_0 + K \dot{\rho}_L^{(1)}(t) \\
&= 2 K \sqrt{[\rho_0 + K \rho_L^{(1)}(t)][\rho_0 + K \rho_R^{(1)}(t)]} \sin{\delta}(t)\\
&\approx \left\{ 2 K \rho_0 + K^2 [ \rho_R^{(1)}(t) + \rho_L^{(1)}(t)] \right\} \sin{\delta}(t).
\end{aligned} \label{B1}
\end{equation}
By comparing the corresponding power of $K$ in Eq. \eqref{B1}, the equation of $\rho_L^{(1)}$ is obtained as
\begin{equation}
\begin{aligned}
\dot{\rho}_L^{(1)} = 2 \rho_0 \sin{\delta(t)}.\\
\end{aligned}\label{approx eq of rho1}
\end{equation}
Similarly, the phase equation of $\rho_R^{(1)}(t)$ can be also obtained as $\dot{\rho}_L^{(1)}(t) = - 2 \rho_0 \sin{\delta(t)}\neq \dot{\rho}_L^{(1)}(t)$ so that $\rho_L(t) \neq \rho_R(t)$. By Eq. \eqref{approx eq of rho1}, the equation of $\phi_L (t)$ is
\begin{equation}
\begin{aligned}
\dot{\phi}_L &\approx -K \sqrt{\frac{\rho_0 + K \rho_R^{(1)}(t)}{\rho_0 + K \rho_L^{(1)}(t)}} \cos{\delta(t)} - \frac{q V}{2}\\
&\approx- K \left\{1 + \frac{K}{2 \rho_0} [\rho_R^{(1)}(t) - \rho_L^{(1)}(t)]\right\} \cos{\delta(t)} - \frac{qV}{2}.
\label{B6}
\end{aligned}
\end{equation}
Similarly, the equation of $\phi_R$ becomes 
\begin{equation}
\dot{\phi}_R(t) \approx - K \left\{1 + \frac{K}{2 \rho_0} [\rho_L^{(1)}(t) - \rho_R^{(1)}(t)]\right\} \cos{\delta(t)} + \frac{qV}{2}.
\label{B7}
\end{equation}
With the help of Eq. \eqref{B6} and \eqref{B7}, the phase difference $\delta(t)$ satisfies
\begin{equation}
\dot{\delta}(t) = - K^2 [\rho_L^{(1)}(t) - \rho_R^{(1)}(t)] + qV \neq qV
\end{equation}
which means the current phase relation of the Josephson effect can only be reliable in the large voltage. Consequently, only the zero order of $K$ in $\rho_\alpha(t)$ is considered in the Feynman approach so that the AC Josephson effect can be obtained. Moreover, the approximation condition of describing the Josephson effect by the Feynman equation is also clarified from the strict solution of Eq. \eqref{equation of Josephson effect}, as shown in the main text.

\twocolumngrid
\bibliography{main}

\begin{thebibliography}{33}%
\makeatletter
\providecommand \@ifxundefined [1]{%
 \@ifx{#1\undefined}
}%
\providecommand \@ifnum [1]{%
 \ifnum #1\expandafter \@firstoftwo
 \else \expandafter \@secondoftwo
 \fi
}%
\providecommand \@ifx [1]{%
 \ifx #1\expandafter \@firstoftwo
 \else \expandafter \@secondoftwo
 \fi
}%
\providecommand \natexlab [1]{#1}%
\providecommand \enquote  [1]{``#1''}%
\providecommand \bibnamefont  [1]{#1}%
\providecommand \bibfnamefont [1]{#1}%
\providecommand \citenamefont [1]{#1}%
\providecommand \href@noop [0]{\@secondoftwo}%
\providecommand \href [0]{\begingroup \@sanitize@url \@href}%
\providecommand \@href[1]{\@@startlink{#1}\@@href}%
\providecommand \@@href[1]{\endgroup#1\@@endlink}%
\providecommand \@sanitize@url [0]{\catcode `\\12\catcode `\$12\catcode `\&12\catcode `\#12\catcode `\^12\catcode `\_12\catcode `\%12\relax}%
\providecommand \@@startlink[1]{}%
\providecommand \@@endlink[0]{}%
\providecommand \url  [0]{\begingroup\@sanitize@url \@url }%
\providecommand \@url [1]{\endgroup\@href {#1}{\urlprefix }}%
\providecommand \urlprefix  [0]{URL }%
\providecommand \Eprint [0]{\href }%
\providecommand \doibase [0]{https://doi.org/}%
\providecommand \selectlanguage [0]{\@gobble}%
\providecommand \bibinfo  [0]{\@secondoftwo}%
\providecommand \bibfield  [0]{\@secondoftwo}%
\providecommand \translation [1]{[#1]}%
\providecommand \BibitemOpen [0]{}%
\providecommand \bibitemStop [0]{}%
\providecommand \bibitemNoStop [0]{.\EOS\space}%
\providecommand \EOS [0]{\spacefactor3000\relax}%
\providecommand \BibitemShut  [1]{\csname bibitem#1\endcsname}%
\let\auto@bib@innerbib\@empty
\bibitem [{\citenamefont {Josephson}(1962)}]{josephson1962possible}%
  \BibitemOpen
  \bibfield  {author} {\bibinfo {author} {\bibfnamefont {B.~D.}\ \bibnamefont {Josephson}},\ }\bibfield  {title} {\bibinfo {title} {Possible new effects in superconductive tunnelling},\ }\href {https://doi.org/10.1016/0031-9163(62)91369-0} {\bibfield  {journal} {\bibinfo  {journal} {Phys. Lett}\ }\textbf {\bibinfo {volume} {1}},\ \bibinfo {pages} {251} (\bibinfo {year} {1962})}\BibitemShut {NoStop}%
\bibitem [{\citenamefont {Anderson}\ and\ \citenamefont {Rowell}(1963)}]{Anderson1963}%
  \BibitemOpen
  \bibfield  {author} {\bibinfo {author} {\bibfnamefont {P.~W.}\ \bibnamefont {Anderson}}\ and\ \bibinfo {author} {\bibfnamefont {J.~M.}\ \bibnamefont {Rowell}},\ }\bibfield  {title} {\bibinfo {title} {Probable observation of the josephson superconducting tunneling effect},\ }\href {https://doi.org/10.1103/PhysRevLett.10.230} {\bibfield  {journal} {\bibinfo  {journal} {Phys. Rev. Lett.}\ }\textbf {\bibinfo {volume} {10}},\ \bibinfo {pages} {230} (\bibinfo {year} {1963})}\BibitemShut {NoStop}%
\bibitem [{\citenamefont {Rowell}(1963)}]{Rowell1963}%
  \BibitemOpen
  \bibfield  {author} {\bibinfo {author} {\bibfnamefont {J.~M.}\ \bibnamefont {Rowell}},\ }\bibfield  {title} {\bibinfo {title} {Magnetic field dependence of the josephson tunnel current},\ }\href {https://doi.org/10.1103/PhysRevLett.11.200} {\bibfield  {journal} {\bibinfo  {journal} {Phys. Rev. Lett.}\ }\textbf {\bibinfo {volume} {11}},\ \bibinfo {pages} {200} (\bibinfo {year} {1963})}\BibitemShut {NoStop}%
\bibitem [{\citenamefont {Shapiro}(1963)}]{Shapiro1963}%
  \BibitemOpen
  \bibfield  {author} {\bibinfo {author} {\bibfnamefont {S.}~\bibnamefont {Shapiro}},\ }\bibfield  {title} {\bibinfo {title} {Josephson currents in superconducting tunneling: The effect of microwaves and other observations},\ }\href {https://doi.org/10.1103/PhysRevLett.11.80} {\bibfield  {journal} {\bibinfo  {journal} {Phys. Rev. Lett.}\ }\textbf {\bibinfo {volume} {11}},\ \bibinfo {pages} {80} (\bibinfo {year} {1963})}\BibitemShut {NoStop}%
\bibitem [{\citenamefont {Ferrell}\ and\ \citenamefont {Prange}(1963)}]{Ferrellmodel}%
  \BibitemOpen
  \bibfield  {author} {\bibinfo {author} {\bibfnamefont {R.~A.}\ \bibnamefont {Ferrell}}\ and\ \bibinfo {author} {\bibfnamefont {R.~E.}\ \bibnamefont {Prange}},\ }\bibfield  {title} {\bibinfo {title} {Self-field limiting of josephson tunneling of superconducting electron pairs},\ }\href {https://link.aps.org/doi/10.1103/PhysRevLett.10.479} {\bibfield  {journal} {\bibinfo  {journal} {Phys. Rev. Lett.}\ }\textbf {\bibinfo {volume} {10}},\ \bibinfo {pages} {479} (\bibinfo {year} {1963})}\BibitemShut {NoStop}%
\bibitem [{\citenamefont {Feynman}\ \emph {et~al.}(2011)\citenamefont {Feynman}, \citenamefont {Leighton},\ and\ \citenamefont {Sands}}]{Feynmanbook}%
  \BibitemOpen
  \bibfield  {author} {\bibinfo {author} {\bibfnamefont {R.~P.}\ \bibnamefont {Feynman}}, \bibinfo {author} {\bibfnamefont {R.~B.}\ \bibnamefont {Leighton}},\ and\ \bibinfo {author} {\bibfnamefont {M.}~\bibnamefont {Sands}},\ }\href@noop {} {\emph {\bibinfo {title} {Feynman Lectures on Physics 3: Quantum Mechanics}}}\ (\bibinfo {year} {2011})\ pp.\ \bibinfo {pages} {602--605}\BibitemShut {NoStop}%
\bibitem [{\citenamefont {Ohta}(1977)}]{Ohtamodel}%
  \BibitemOpen
  \bibfield  {author} {\bibinfo {author} {\bibfnamefont {H.}~\bibnamefont {Ohta}},\ }\bibfield  {title} {\bibinfo {title} {A self-consistent model of the josephson junction},\ }\href {https://www.degruyter.com/document/doi/10.1515/9783110887495-005/html#MLA} {\bibfield  {journal} {\bibinfo  {journal} {IC SQUID}\ }\textbf {\bibinfo {volume} {76}},\ \bibinfo {pages} {35} (\bibinfo {year} {1977})}\BibitemShut {NoStop}%
\bibitem [{\citenamefont {Bassani}\ \emph {et~al.}(2005)\citenamefont {Bassani}, \citenamefont {Liedl},\ and\ \citenamefont {Wyder}}]{Bassani2005EncyclopediaOC}%
  \BibitemOpen
  \bibfield  {author} {\bibinfo {author} {\bibfnamefont {G.~F.}\ \bibnamefont {Bassani}}, \bibinfo {author} {\bibfnamefont {G.~L.}\ \bibnamefont {Liedl}},\ and\ \bibinfo {author} {\bibfnamefont {P.}~\bibnamefont {Wyder}},\ }\bibfield  {title} {\bibinfo {title} {Encyclopedia of condensed matter physics}\ }(\bibinfo {year} {2005})\ pp.\ \bibinfo {pages} {110--111}\BibitemShut {NoStop}%
\bibitem [{\citenamefont {Poole}\ \emph {et~al.}(2014)\citenamefont {Poole}, \citenamefont {Farach}, \citenamefont {Creswick},\ and\ \citenamefont {Prozorov}}]{Superconductivity}%
  \BibitemOpen
  \bibfield  {author} {\bibinfo {author} {\bibfnamefont {C.~P.}\ \bibnamefont {Poole}}, \bibinfo {author} {\bibfnamefont {H.~A.}\ \bibnamefont {Farach}}, \bibinfo {author} {\bibfnamefont {R.~J.}\ \bibnamefont {Creswick}},\ and\ \bibinfo {author} {\bibfnamefont {R.}~\bibnamefont {Prozorov}},\ }\href@noop {} {\emph {\bibinfo {title} {Superconductivity}}}\ (\bibinfo {year} {2014})\ pp.\ \bibinfo {pages} {537--539}\BibitemShut {NoStop}%
\bibitem [{\citenamefont {De~Luca}(2020)}]{Roberto}%
  \BibitemOpen
  \bibfield  {author} {\bibinfo {author} {\bibfnamefont {R.}~\bibnamefont {De~Luca}},\ }\bibfield  {title} {\bibinfo {title} {Magnetic properties of josephson junction networks}\ }(\bibinfo {year} {2020})\ pp.\ \bibinfo {pages} {3--5}\BibitemShut {NoStop}%
\bibitem [{\citenamefont {Ginzburg}\ and\ \citenamefont {Landau}(1950)}]{GL}%
  \BibitemOpen
  \bibfield  {author} {\bibinfo {author} {\bibfnamefont {V.~L.}\ \bibnamefont {Ginzburg}}\ and\ \bibinfo {author} {\bibfnamefont {L.~D.}\ \bibnamefont {Landau}},\ }\bibfield  {title} {\bibinfo {title} {{On the Theory of superconductivity}},\ }\href {https://doi.org/10.1016/B978-0-08-010586-4.50035-3} {\bibfield  {journal} {\bibinfo  {journal} {Zh. Eksp. Teor. Fiz.}\ }\textbf {\bibinfo {volume} {20}},\ \bibinfo {pages} {1064} (\bibinfo {year} {1950})}\BibitemShut {NoStop}%
\bibitem [{\citenamefont {Gor’kov}(1959)}]{gor1959microscopic}%
  \BibitemOpen
  \bibfield  {author} {\bibinfo {author} {\bibfnamefont {L.~P.}\ \bibnamefont {Gor’kov}},\ }\bibfield  {title} {\bibinfo {title} {Microscopic derivation of the ginzburg-landau equations in the theory of superconductivity},\ }\href {http://www.jetp.ras.ru/cgi-bin/dn/e_009_06_1364.pdf} {\bibfield  {journal} {\bibinfo  {journal} {Sov. Phys. JETP}\ }\textbf {\bibinfo {volume} {9}},\ \bibinfo {pages} {1364} (\bibinfo {year} {1959})}\BibitemShut {NoStop}%
\bibitem [{\citenamefont {Coullet}\ \emph {et~al.}(1985)\citenamefont {Coullet}, \citenamefont {Fauve},\ and\ \citenamefont {Tirapegui}}]{Coullet1985LargeSI}%
  \BibitemOpen
  \bibfield  {author} {\bibinfo {author} {\bibfnamefont {P.}~\bibnamefont {Coullet}}, \bibinfo {author} {\bibfnamefont {S.}~\bibnamefont {Fauve}},\ and\ \bibinfo {author} {\bibfnamefont {E.}~\bibnamefont {Tirapegui}},\ }\bibfield  {title} {\bibinfo {title} {Large scale instability of nonlinear standing waves},\ }\href {https://api.semanticscholar.org/CorpusID:32208819} {\bibfield  {journal} {\bibinfo  {journal} {J. Physique Lett.}\ }\textbf {\bibinfo {volume} {46}},\ \bibinfo {pages} {787} (\bibinfo {year} {1985})}\BibitemShut {NoStop}%
\bibitem [{\citenamefont {Coullet}\ \emph {et~al.}(1987)\citenamefont {Coullet}, \citenamefont {Elphick}, \citenamefont {Gil},\ and\ \citenamefont {Lega}}]{coullet1987topological}%
  \BibitemOpen
  \bibfield  {author} {\bibinfo {author} {\bibfnamefont {P.}~\bibnamefont {Coullet}}, \bibinfo {author} {\bibfnamefont {C.}~\bibnamefont {Elphick}}, \bibinfo {author} {\bibfnamefont {L.}~\bibnamefont {Gil}},\ and\ \bibinfo {author} {\bibfnamefont {J.}~\bibnamefont {Lega}},\ }\bibfield  {title} {\bibinfo {title} {Topological defects of wave patterns},\ }\href {https://journals.aps.org/prl/abstract/10.1103/PhysRevLett.59.884} {\bibfield  {journal} {\bibinfo  {journal} {Phys. Rev. Lett.}\ }\textbf {\bibinfo {volume} {59}},\ \bibinfo {pages} {884} (\bibinfo {year} {1987})}\BibitemShut {NoStop}%
\bibitem [{\citenamefont {Ledesma-Dur{\'a}n}\ and\ \citenamefont {Arag{\'o}n}(2020)}]{GLS1}%
  \BibitemOpen
  \bibfield  {author} {\bibinfo {author} {\bibfnamefont {A.}~\bibnamefont {Ledesma-Dur{\'a}n}}\ and\ \bibinfo {author} {\bibfnamefont {J.~L.}\ \bibnamefont {Arag{\'o}n}},\ }\bibfield  {title} {\bibinfo {title} {Spatio-temporal numerical solutions of the coupled real and complex ginzburg-landau amplitude equations for one-dimensional systems near the turing-hopf bifurcation},\ }\href {https://www.sciencedirect.com/science/article/abs/pii/S1007570419304642} {\bibfield  {journal} {\bibinfo  {journal} {Commun. Nonlinear Sci. Numer. Simul.}\ }\textbf {\bibinfo {volume} {83}},\ \bibinfo {pages} {105145} (\bibinfo {year} {2020})}\BibitemShut {NoStop}%
\bibitem [{\citenamefont {Smerzi}\ \emph {et~al.}(1997)\citenamefont {Smerzi}, \citenamefont {Fantoni}, \citenamefont {Giovanazzi},\ and\ \citenamefont {Shenoy}}]{smerzi1997quantum}%
  \BibitemOpen
  \bibfield  {author} {\bibinfo {author} {\bibfnamefont {A.}~\bibnamefont {Smerzi}}, \bibinfo {author} {\bibfnamefont {S.}~\bibnamefont {Fantoni}}, \bibinfo {author} {\bibfnamefont {S.}~\bibnamefont {Giovanazzi}},\ and\ \bibinfo {author} {\bibfnamefont {S.}~\bibnamefont {Shenoy}},\ }\bibfield  {title} {\bibinfo {title} {Quantum coherent atomic tunneling between two trapped bose-einstein condensates},\ }\href {https://journals.aps.org/prl/abstract/10.1103/PhysRevLett.79.4950} {\bibfield  {journal} {\bibinfo  {journal} {Phys. Rev. Lett.}\ }\textbf {\bibinfo {volume} {79}},\ \bibinfo {pages} {4950} (\bibinfo {year} {1997})}\BibitemShut {NoStop}%
\bibitem [{\citenamefont {Radzihovsky}\ and\ \citenamefont {Gurarie}(2010)}]{radzihovsky2010relation}%
  \BibitemOpen
  \bibfield  {author} {\bibinfo {author} {\bibfnamefont {L.}~\bibnamefont {Radzihovsky}}\ and\ \bibinfo {author} {\bibfnamefont {V.}~\bibnamefont {Gurarie}},\ }\bibfield  {title} {\bibinfo {title} {Relation between ac josephson effect and double-well bose-einstein-condensate oscillations},\ }\href {https://journals.aps.org/pra/abstract/10.1103/PhysRevA.81.063609} {\bibfield  {journal} {\bibinfo  {journal} {Phys. Rev. A}\ }\textbf {\bibinfo {volume} {81}},\ \bibinfo {pages} {063609} (\bibinfo {year} {2010})}\BibitemShut {NoStop}%
\bibitem [{\citenamefont {Piselli}\ \emph {et~al.}(2020)\citenamefont {Piselli}, \citenamefont {Simonucci},\ and\ \citenamefont {Strinati}}]{piselli2020josephson}%
  \BibitemOpen
  \bibfield  {author} {\bibinfo {author} {\bibfnamefont {V.}~\bibnamefont {Piselli}}, \bibinfo {author} {\bibfnamefont {S.}~\bibnamefont {Simonucci}},\ and\ \bibinfo {author} {\bibfnamefont {G.~C.}\ \bibnamefont {Strinati}},\ }\bibfield  {title} {\bibinfo {title} {Josephson effect at finite temperature along the bcs-bec crossover},\ }\href {https://journals.aps.org/prb/abstract/10.1103/PhysRevB.102.144517} {\bibfield  {journal} {\bibinfo  {journal} {Phys. Rev. B}\ }\textbf {\bibinfo {volume} {102}},\ \bibinfo {pages} {144517} (\bibinfo {year} {2020})}\BibitemShut {NoStop}%
\bibitem [{\citenamefont {Levy}\ \emph {et~al.}(2007)\citenamefont {Levy}, \citenamefont {Lahoud}, \citenamefont {Shomroni},\ and\ \citenamefont {Steinhauer}}]{levy2007ac}%
  \BibitemOpen
  \bibfield  {author} {\bibinfo {author} {\bibfnamefont {S.}~\bibnamefont {Levy}}, \bibinfo {author} {\bibfnamefont {E.}~\bibnamefont {Lahoud}}, \bibinfo {author} {\bibfnamefont {I.}~\bibnamefont {Shomroni}},\ and\ \bibinfo {author} {\bibfnamefont {J.}~\bibnamefont {Steinhauer}},\ }\bibfield  {title} {\bibinfo {title} {The ac and dc josephson effects in a bose-einstein condensate},\ }\href {https://www.nature.com/articles/nature06186} {\bibfield  {journal} {\bibinfo  {journal} {Nature}\ }\textbf {\bibinfo {volume} {449}},\ \bibinfo {pages} {579} (\bibinfo {year} {2007})}\BibitemShut {NoStop}%
\bibitem [{\citenamefont {De~Luca}\ and\ \citenamefont {Romeo}(2009)}]{DeLuca}%
  \BibitemOpen
  \bibfield  {author} {\bibinfo {author} {\bibfnamefont {R.}~\bibnamefont {De~Luca}}\ and\ \bibinfo {author} {\bibfnamefont {F.}~\bibnamefont {Romeo}},\ }\bibfield  {title} {\bibinfo {title} {Sawtooth current-phase relation of a superconducting trilayer system described using ohta's formalism},\ }\href {https://link.aps.org/doi/10.1103/PhysRevB.79.094516} {\bibfield  {journal} {\bibinfo  {journal} {Phys. Rev. B}\ }\textbf {\bibinfo {volume} {79}},\ \bibinfo {pages} {094516} (\bibinfo {year} {2009})}\BibitemShut {NoStop}%
\bibitem [{\citenamefont {Ouassou}\ and\ \citenamefont {Linder}(2017)}]{ouassou2017spin}%
  \BibitemOpen
  \bibfield  {author} {\bibinfo {author} {\bibfnamefont {J.~A.}\ \bibnamefont {Ouassou}}\ and\ \bibinfo {author} {\bibfnamefont {J.}~\bibnamefont {Linder}},\ }\bibfield  {title} {\bibinfo {title} {Spin-switch josephson junctions with magnetically tunable $\sin (\delta \varphi/n)$ current-phase relation},\ }\href {https://journals.aps.org/prb/abstract/10.1103/PhysRevB.96.064516} {\bibfield  {journal} {\bibinfo  {journal} {Phys. Rev. B}\ }\textbf {\bibinfo {volume} {96}},\ \bibinfo {pages} {064516} (\bibinfo {year} {2017})}\BibitemShut {NoStop}%
\bibitem [{\citenamefont {Barash}(2018)}]{barash2018proximity}%
  \BibitemOpen
  \bibfield  {author} {\bibinfo {author} {\bibfnamefont {Y.~S.}\ \bibnamefont {Barash}},\ }\bibfield  {title} {\bibinfo {title} {Proximity-reduced range of internal phase differences in double josephson junctions with closely spaced interfaces},\ }\href {https://journals.aps.org/prb/abstract/10.1103/PhysRevB.97.224509} {\bibfield  {journal} {\bibinfo  {journal} {Phys. Rev. B}\ }\textbf {\bibinfo {volume} {97}},\ \bibinfo {pages} {224509} (\bibinfo {year} {2018})}\BibitemShut {NoStop}%
\bibitem [{\citenamefont {Tinkham}(2004)}]{tinkham2004introduction}%
  \BibitemOpen
  \bibfield  {author} {\bibinfo {author} {\bibfnamefont {M.}~\bibnamefont {Tinkham}},\ }\href@noop {} {\emph {\bibinfo {title} {Introduction to superconductivity}}}\ (\bibinfo {year} {2004})\ pp.\ \bibinfo {pages} {274--276}\BibitemShut {NoStop}%
\bibitem [{\citenamefont {Banszerus}\ \emph {et~al.}(2024)\citenamefont {Banszerus}, \citenamefont {Marshall}, \citenamefont {Andersson}, \citenamefont {Lindemann}, \citenamefont {Manfra}, \citenamefont {Marcus},\ and\ \citenamefont {Vaitiek{\.e}nas}}]{banszerus2024voltage}%
  \BibitemOpen
  \bibfield  {author} {\bibinfo {author} {\bibfnamefont {L.}~\bibnamefont {Banszerus}}, \bibinfo {author} {\bibfnamefont {W.}~\bibnamefont {Marshall}}, \bibinfo {author} {\bibfnamefont {C.}~\bibnamefont {Andersson}}, \bibinfo {author} {\bibfnamefont {T.}~\bibnamefont {Lindemann}}, \bibinfo {author} {\bibfnamefont {M.}~\bibnamefont {Manfra}}, \bibinfo {author} {\bibfnamefont {C.}~\bibnamefont {Marcus}},\ and\ \bibinfo {author} {\bibfnamefont {S.}~\bibnamefont {Vaitiek{\.e}nas}},\ }\bibfield  {title} {\bibinfo {title} {Voltage-controlled synthesis of higher harmonics in hybrid josephson junction circuits},\ }\href {https://arxiv.org/abs/2402.11603} {\bibfield  {journal} {\bibinfo  {journal} {arXiv:2402.11603}\ } (\bibinfo {year} {2024})}\BibitemShut {NoStop}%
\bibitem [{\citenamefont {Bozkurt}\ \emph {et~al.}(2023)\citenamefont {Bozkurt}, \citenamefont {Brookman}, \citenamefont {Fatemi},\ and\ \citenamefont {Akhmerov}}]{bozkurt2023double}%
  \BibitemOpen
  \bibfield  {author} {\bibinfo {author} {\bibfnamefont {A.~M.}\ \bibnamefont {Bozkurt}}, \bibinfo {author} {\bibfnamefont {J.}~\bibnamefont {Brookman}}, \bibinfo {author} {\bibfnamefont {V.}~\bibnamefont {Fatemi}},\ and\ \bibinfo {author} {\bibfnamefont {A.~R.}\ \bibnamefont {Akhmerov}},\ }\bibfield  {title} {\bibinfo {title} {Double-fourier engineering of josephson energy-phase relationships applied to diodes},\ }\href {https://scipost.org/SciPostPhys.15.5.204} {\bibfield  {journal} {\bibinfo  {journal} {SciPost Phys.}\ }\textbf {\bibinfo {volume} {15}},\ \bibinfo {pages} {204} (\bibinfo {year} {2023})}\BibitemShut {NoStop}%
\bibitem [{\citenamefont {Wallace}\ and\ \citenamefont {Stavn}(1965)}]{Wallace1965}%
  \BibitemOpen
  \bibfield  {author} {\bibinfo {author} {\bibfnamefont {P.}~\bibnamefont {Wallace}}\ and\ \bibinfo {author} {\bibfnamefont {M.}~\bibnamefont {Stavn}},\ }\bibfield  {title} {\bibinfo {title} {Quasi-spin treatment of josephson tunneling between superconductors},\ }\href {https://doi.org/10.1139/p65-037} {\bibfield  {journal} {\bibinfo  {journal} {Can. J. Phys}\ }\textbf {\bibinfo {volume} {43}},\ \bibinfo {pages} {411} (\bibinfo {year} {1965})}\BibitemShut {NoStop}%
\bibitem [{\citenamefont {Lee}\ and\ \citenamefont {Scully}(1971)}]{lee1971theory}%
  \BibitemOpen
  \bibfield  {author} {\bibinfo {author} {\bibfnamefont {P.~A.}\ \bibnamefont {Lee}}\ and\ \bibinfo {author} {\bibfnamefont {M.~O.}\ \bibnamefont {Scully}},\ }\bibfield  {title} {\bibinfo {title} {Theory of josephson radiation. i. general theory},\ }\href {https://journals.aps.org/prb/abstract/10.1103/PhysRevB.3.769} {\bibfield  {journal} {\bibinfo  {journal} {Phys. Rev. B}\ }\textbf {\bibinfo {volume} {3}},\ \bibinfo {pages} {769} (\bibinfo {year} {1971})}\BibitemShut {NoStop}%
\bibitem [{\citenamefont {Gupta}\ \emph {et~al.}(2023)\citenamefont {Gupta}, \citenamefont {Graziano}, \citenamefont {Pendharkar}, \citenamefont {Dong}, \citenamefont {Dempsey}, \citenamefont {Palmstr{\o}m},\ and\ \citenamefont {Pribiag}}]{gupta2023gate}%
  \BibitemOpen
  \bibfield  {author} {\bibinfo {author} {\bibfnamefont {M.}~\bibnamefont {Gupta}}, \bibinfo {author} {\bibfnamefont {G.~V.}\ \bibnamefont {Graziano}}, \bibinfo {author} {\bibfnamefont {M.}~\bibnamefont {Pendharkar}}, \bibinfo {author} {\bibfnamefont {J.~T.}\ \bibnamefont {Dong}}, \bibinfo {author} {\bibfnamefont {C.~P.}\ \bibnamefont {Dempsey}}, \bibinfo {author} {\bibfnamefont {C.}~\bibnamefont {Palmstr{\o}m}},\ and\ \bibinfo {author} {\bibfnamefont {V.~S.}\ \bibnamefont {Pribiag}},\ }\bibfield  {title} {\bibinfo {title} {Gate-tunable superconducting diode effect in a three-terminal josephson device},\ }\href {https://www.nature.com/articles/s41467-023-38856-0} {\bibfield  {journal} {\bibinfo  {journal} {Nat. Commun}\ }\textbf {\bibinfo {volume} {14}},\ \bibinfo {pages} {3078} (\bibinfo {year} {2023})}\BibitemShut {NoStop}%
\bibitem [{\citenamefont {Hu}\ \emph {et~al.}(2007)\citenamefont {Hu}, \citenamefont {Wu},\ and\ \citenamefont {Dai}}]{hu2007proposed}%
  \BibitemOpen
  \bibfield  {author} {\bibinfo {author} {\bibfnamefont {J.}~\bibnamefont {Hu}}, \bibinfo {author} {\bibfnamefont {C.}~\bibnamefont {Wu}},\ and\ \bibinfo {author} {\bibfnamefont {X.}~\bibnamefont {Dai}},\ }\bibfield  {title} {\bibinfo {title} {Proposed design of a josephson diode},\ }\href {https://journals.aps.org/prl/abstract/10.1103/PhysRevLett.99.067004} {\bibfield  {journal} {\bibinfo  {journal} {Phys. Rev. Lett.}\ }\textbf {\bibinfo {volume} {99}},\ \bibinfo {pages} {067004} (\bibinfo {year} {2007})}\BibitemShut {NoStop}%
\bibitem [{\citenamefont {Wu}\ \emph {et~al.}(2022)\citenamefont {Wu}, \citenamefont {Wang}, \citenamefont {Xu}, \citenamefont {Sivakumar}, \citenamefont {Pasco}, \citenamefont {Filippozzi}, \citenamefont {Parkin}, \citenamefont {Zeng}, \citenamefont {McQueen},\ and\ \citenamefont {Ali}}]{wu2022field}%
  \BibitemOpen
  \bibfield  {author} {\bibinfo {author} {\bibfnamefont {H.}~\bibnamefont {Wu}}, \bibinfo {author} {\bibfnamefont {Y.}~\bibnamefont {Wang}}, \bibinfo {author} {\bibfnamefont {Y.}~\bibnamefont {Xu}}, \bibinfo {author} {\bibfnamefont {P.~K.}\ \bibnamefont {Sivakumar}}, \bibinfo {author} {\bibfnamefont {C.}~\bibnamefont {Pasco}}, \bibinfo {author} {\bibfnamefont {U.}~\bibnamefont {Filippozzi}}, \bibinfo {author} {\bibfnamefont {S.~S.}\ \bibnamefont {Parkin}}, \bibinfo {author} {\bibfnamefont {Y.-J.}\ \bibnamefont {Zeng}}, \bibinfo {author} {\bibfnamefont {T.}~\bibnamefont {McQueen}},\ and\ \bibinfo {author} {\bibfnamefont {M.~N.}\ \bibnamefont {Ali}},\ }\bibfield  {title} {\bibinfo {title} {The field-free josephson diode in a van der waals heterostructure},\ }\href {https://www.nature.com/articles/s41586-022-04504-8} {\bibfield  {journal} {\bibinfo  {journal} {Nature}\ }\textbf {\bibinfo {volume} {604}},\ \bibinfo {pages} {653} (\bibinfo {year} {2022})}\BibitemShut {NoStop}%
\bibitem [{\citenamefont {Souto}\ \emph {et~al.}(2022)\citenamefont {Souto}, \citenamefont {Leijnse},\ and\ \citenamefont {Schrade}}]{souto2022josephson}%
  \BibitemOpen
  \bibfield  {author} {\bibinfo {author} {\bibfnamefont {R.~S.}\ \bibnamefont {Souto}}, \bibinfo {author} {\bibfnamefont {M.}~\bibnamefont {Leijnse}},\ and\ \bibinfo {author} {\bibfnamefont {C.}~\bibnamefont {Schrade}},\ }\bibfield  {title} {\bibinfo {title} {Josephson diode effect in supercurrent interferometers},\ }\href {https://journals.aps.org/prl/abstract/10.1103/PhysRevLett.129.267702} {\bibfield  {journal} {\bibinfo  {journal} {Phys. Rev. Lett.}\ }\textbf {\bibinfo {volume} {129}},\ \bibinfo {pages} {267702} (\bibinfo {year} {2022})}\BibitemShut {NoStop}%
\bibitem [{\citenamefont {Gutfreund}\ \emph {et~al.}(2023)\citenamefont {Gutfreund}, \citenamefont {Matsuki}, \citenamefont {Plastovets}, \citenamefont {Noah}, \citenamefont {Gorzawski}, \citenamefont {Fridman}, \citenamefont {Yang}, \citenamefont {Buzdin}, \citenamefont {Millo}, \citenamefont {Robinson} \emph {et~al.}}]{gutfreund2023direct}%
  \BibitemOpen
  \bibfield  {author} {\bibinfo {author} {\bibfnamefont {A.}~\bibnamefont {Gutfreund}}, \bibinfo {author} {\bibfnamefont {H.}~\bibnamefont {Matsuki}}, \bibinfo {author} {\bibfnamefont {V.}~\bibnamefont {Plastovets}}, \bibinfo {author} {\bibfnamefont {A.}~\bibnamefont {Noah}}, \bibinfo {author} {\bibfnamefont {L.}~\bibnamefont {Gorzawski}}, \bibinfo {author} {\bibfnamefont {N.}~\bibnamefont {Fridman}}, \bibinfo {author} {\bibfnamefont {G.}~\bibnamefont {Yang}}, \bibinfo {author} {\bibfnamefont {A.}~\bibnamefont {Buzdin}}, \bibinfo {author} {\bibfnamefont {O.}~\bibnamefont {Millo}}, \bibinfo {author} {\bibfnamefont {J.~W.}\ \bibnamefont {Robinson}}, \emph {et~al.},\ }\bibfield  {title} {\bibinfo {title} {Direct observation of a superconducting vortex diode},\ }\href {https://www.nature.com/articles/s41467-023-37294-2} {\bibfield  {journal} {\bibinfo  {journal} {Nat. Commun.}\ }\textbf {\bibinfo {volume} {14}},\ \bibinfo {pages} {1630} (\bibinfo {year} {2023})}\BibitemShut {NoStop}%
\bibitem [{\citenamefont {Romeo}\ and\ \citenamefont {De~Luca}(2022)}]{romeo2022cooper}%
  \BibitemOpen
  \bibfield  {author} {\bibinfo {author} {\bibfnamefont {F.}~\bibnamefont {Romeo}}\ and\ \bibinfo {author} {\bibfnamefont {R.}~\bibnamefont {De~Luca}},\ }\bibfield  {title} {\bibinfo {title} {Cooper pairs localization in tree-like networks of superconducting islands},\ }\href {https://link.springer.com/article/10.1140/epjp/s13360-022-02928-9} {\bibfield  {journal} {\bibinfo  {journal} {Eur. Phys. J. Plus.}\ }\textbf {\bibinfo {volume} {137}},\ \bibinfo {pages} {726} (\bibinfo {year} {2022})}\BibitemShut {NoStop}%
\end{thebibliography}%
\end{document}